\documentclass[12pt]{article}
\usepackage{latexsym}
\usepackage[cp1250]{inputenc}

\title{On the unitarity of higher-derivative and nonlocal theories}
\author{Katarzyna Bolonek\thanks{supported by the grant 1 P03B 125 29 of the Polish Ministry of Science.},\\ 
 Piotr  Kosi\'nski\thanks{supported by the grant 1 P03B 021 28 of the Polish Ministry of Science.} \\
Department of Theoretical Physics II \\
University of {\L}\'od\'z \\
Pomorska 149/153, 90 - 236 {\L}\'od\'z, Poland.}
\date{}

\begin{document}
\maketitle
\begin{abstract}
We consider two simple models of higher-derivative and nonlocal quantum systems. It is shown that, contrary to some claims found in 
literature, they can be made unitary
\end{abstract}

\newpage
Considerable attention has been paid recently to the problem of quantizing the theories which contain higher derivatives or are even nonlocal in time. 
This is mainly due to the fact that such theories arise naturally when one attempts to model the dynamics on noncommutative space-time with the help 
of star product construction \cite{b1} $\div $\ \cite{b3}. It has been shown that the field theories with space-time noncommutativity lead to 
nonunitarity and acausality, at least when quantized with the help of naive Feynman rules \cite{b4} $\div $\ \cite{b6}. There are several proposals 
of alternative quantization schemes which seem to cure the trouble with nonunitarity \cite{b7} $\div $\ \cite{b10}. However, they give rise to other 
problems \cite{b11}, \cite{b12} so the question of the existence of consistent quantum theory remains open. In view of that one is tempted to study 
simple models to gain more understanding of the problems we are faced with. However, even in the simplest cases there is some misunderstanding 
concerning the consistency of the relevant models. In order to clarify some issues we study two simple systems which are sometimes claimed 
to lead to nonunitary evolution. 

Let us first consider the Lagrangian \cite{b13} 
\begin{eqnarray}
L=\frac{1}{2}\left(\ddot q^2-\Omega ^4q^2\right)   \label{w1}
\end{eqnarray}
The corresponding classical equation of motion reads
\begin{eqnarray}
q^{(IV)}-\Omega ^4q=0  \label{w2}
\end{eqnarray}
The general solution to eq.(\ref{w2}) is a linear combination of $exp(\pm i\Omega t)$\ and $exp(\pm \Omega t)$. It is sometimes claimed \cite{b13} that 
due to the existence of exponentially rising solutions $exp(\pm \Omega t)$\ the quantum Hamiltonian of such type of systems is not  hermitean   and the evolution 
operator $exp\left(\frac{-itH}{\hbar }\right)$\ is not unitary. 

Let us consider this problem in some detail. In order to quantize the model we have to put it first into Hamiltonian form. This can be done with the use of 
Ostrogradsky formalism \cite{b14} $\div $\ \cite{b16}. One defines the canonical variables
\begin{eqnarray}
&& q_1\equiv q, \,\,\, q_2\equiv \dot q   \nonumber \\
&& \Pi _1\equiv \frac{\delta L}{\delta \dot q}=\frac{\partial L}{\partial \dot q}-\frac{d}{dt}\left(\frac{\partial L}{\partial \ddot q}\right)=-
\stackrel{\ldots}{q}   \label{w3}  \\
&& \Pi _2\equiv \frac{\delta L}{\delta \ddot q}=\frac{\partial L}{\partial \ddot q}=\ddot q  \nonumber
\end{eqnarray}
and the Hamiltonian
\begin{eqnarray}
H\equiv \Pi _1\dot q_1+\Pi _2\dot q_2-L=\Pi _1q_2+\frac{1}{2}\Pi _2^2+\frac{\Omega ^4}{2}q_1^2  \label{w4}
\end{eqnarray}
It is straightforward to check that the canonical equations of motion imply the initial equation for $q$. \\
Our system, being linear, can be immediately quantized. The question we intend to solve is whether the resulting quantum system is unitary. The argument 
against unitarity based on the existence of exponentially growing solutions to classical equations of motion might go as follows. Let
\begin{eqnarray}
\hat X\equiv \hat \Pi _1-\Omega \hat \Pi _2-\Omega ^3\hat q_1-\Omega ^2\hat q_2  \label{w5}
\end{eqnarray}
then
\begin{eqnarray}
[\hat X,\hat H]=i\hbar \Omega \hat X  \label{w6}
\end{eqnarray}
(which is equivalent to the existence of exponentially growing solutions to Heisenberg equations of motion). Assume now that $\hat H$\ is selfadjoint 
and let $\mid E\rangle$\ be the eigenvectors of $\hat H$\ (possibly the generalized ones, corresponding to continous spectrum). Then eq. (\ref{w6}) 
implies
\begin{eqnarray}
(E-E'-i\hbar \Omega )\langle E'\mid \hat X\mid E \rangle =0     \label{w7}
\end{eqnarray}
Eq. (\ref{w7}) is fulfiled only provided
\begin{eqnarray}
\langle E'\mid \hat X\mid E\rangle =0 \label{w8}
\end{eqnarray}
for all $E,E'$\ which in turn means that the canonical variables are lineary dependent contradicting the canonical commutation rules. 

The only weak point in the above reasoning is the assumption on the existence (even in distributional sense) of the matrix elements
$\langle E'\mid \hat X\mid E \rangle $. In fact, we shall show below that the theory is unitary while $\langle E'\mid \hat X\mid E \rangle $\ are 
not well defined. \\
To this end we make a simple canonical (hence unitary) transformation
\begin{eqnarray}
&& \hat q_1=\frac{1}{\sqrt{2}\Omega }\left(\hat X_1+\hat X_2\right), \;\;\;\; \hat q_2=\frac{1}{\sqrt{2}\Omega }\left(\hat P_1-\hat P_2\right) \nonumber \\
&& \Pi _1=\frac{\Omega }{\sqrt{2}}\left(\hat P_1+\hat P_2\right), \;\;\;\;\;\; \hat \Pi _2=\frac{\Omega }{\sqrt{2}}\left(-\hat X_1+\hat X_2\right)  \label{w9}
\end{eqnarray}
Then the Hamiltonian takes the following form
\begin{eqnarray}
\hat H=\left(\frac{1}{2}\hat P_1^2+\frac{\Omega }{2}\hat X_1^2\right)-\left(\frac{1}{2}\hat P_2^2-\frac{\Omega }{2}\hat X_2^2\right)   \label{w10}
\end{eqnarray}
$\hat H$\ is the difference of harmonic oscillator and the inverted harmonic oscillator. Both terms depend on different variables so if they are selfadjoint 
and generate unitary dynamics so is $\hat H$. Only the second piece calls for some explanation. It is unbounded from below but only quadratically. Therefore, 
it defines selfadjoint operator with purely continuous spectrum covering the whole real axis \cite{b17}. 

In order to find the coordinate representation for eigenvectors of $\hat H$\ we have only to find the corresponding wave functions for inverted harmonic 
oscilator. This is quite straightforward \cite{b18} and results in the following final formula for eigenfunctions of $\hat H$.
\begin{eqnarray}
&& \langle X_1,X_2\mid n,\varepsilon ;\pm \rangle=  \label{w11} \\
&& =\left(\frac{1}{\sqrt{\sqrt{\pi }2^nn!}}
\sqrt[4]{\frac{\Omega }{\hbar }}H_n \nonumber   
 \left(\sqrt{\frac{\Omega }{\hbar }}X_1\right)e^{-\frac{\Omega X_1^2}
{2\hbar }}\right) \nonumber  \\
&&\left(\sqrt[4]{\frac{2\Omega }{\hbar }}\frac{e^{\frac{\varepsilon \pi }{4\hbar \Omega }}}{\sqrt{4\pi ch(\frac{\pi \varepsilon }
{\hbar \Omega })}}D_{-\frac{i\varepsilon }{\hbar \Omega }-\frac{1}{2}}\left(\pm \sqrt{\frac{2\Omega }{\hbar }}\left(\frac{1+i}{\sqrt{2}}\right)X_2
\right)\right);  \nonumber 
\end{eqnarray}
here $H_n$\ are Hermite polynomials while $D_\nu $\ - the parabolic cylinder functions. The energies of the above states are given by 
\begin{eqnarray}
E\equiv E(n,\varepsilon )=\hbar \Omega \left(n+\frac{1}{2}\right)-\varepsilon   \label{w12}
\end{eqnarray}
Consider now the matrix elements $\langle E'\mid \hat X\mid E \rangle $. By virtue of eqs. (\ref{w5}) and (\ref{w9}) we find
\begin{eqnarray}
\hat X=\sqrt{2}\Omega (\hat P_2-\Omega \hat X_2)  \label{w13}
\end{eqnarray}
Now, using (\ref{w11}), (\ref{w13}) and the asymptotic form of parabolic cylinder functions \cite{b18} we find that the matrix elements 
$\langle n',\varepsilon ',\lambda '\mid \hat X\mid n,\varepsilon ,\lambda  \rangle , \;\;\; \lambda ,\lambda '=\pm 1$, are expressed, apart from regular 
contributions, in terms of badly divergent integrals of the form
\begin{eqnarray}
\int\limits^\infty dxe^{\frac{i(\varepsilon '-\varepsilon )}{\hbar \Omega } ln x}\sim \int\limits^\infty dy e^{y\left(1+i\frac{(\varepsilon -\varepsilon ')}
{\hbar \Omega }\right)}   \label{w14}
\end{eqnarray}
Therefore, the matrix elements under consideration are not well defined even in the distributional sense. 

As a second model we consider the nonlocal harmonic oscillator \cite{b19}, \cite{b20}
\begin{eqnarray}
L=\frac{1}{2}\dot q(t)^2-\frac{\omega ^2}{2}q{\left(t-\frac{T}{2}\right)}q{\left(t+\frac{T}{2}\right)}  \label{w15}
\end{eqnarray}
Euler-Lagrange equation
\begin{eqnarray}
\int dt'\frac{\delta L(t')}{\delta q(t)}=0  \label{w16}
\end{eqnarray}
reads
\begin{eqnarray}
\ddot q(t)+\frac{\omega ^2}{2}\left(q(t-T)+q(t+T)\right)=0  \label{w17}
\end{eqnarray}
This theory can be quantized using the method of Pais and Uhlenbeck \cite{b21}. To this end we rewrite the Lagrangian in equivalent form
\begin{eqnarray}
L=-\frac{1}{2}q(t)\ddot q(t)-\frac{\omega ^2}{4}((q(t)q(t-T)+q(t)q(t+T))   \label{w18}
\end{eqnarray}
or
\begin{eqnarray}
L=-\frac{1}{2}q(t)\left( \left(\frac{d}{dt}\right)^2+\omega ^2ch\left(T\frac{d}{dt}\right)\right)q(t)  \label{w19}
\end{eqnarray}      
According to the prescription given in Ref. \cite{b21} we consider the entire function
\begin{eqnarray}
\Phi (u)\equiv z^2+\omega ^2ch(Tz), \;\;\; u=z^2    \label{w20}
\end{eqnarray}
It is not difficult to verify that: $(i) \;\Phi (u)$\ has no multiple zeros except a discrete set of values of $\omega T$; in the latter case the double 
zeros are real and negative, $(ii)$\ there is a finite nonempty set of real negative zeros, $z^2=-\Omega ^2_i, \;\; i=1,...,m, \;\; m\geq 1, \;\;(iii)$\ 
there is an infinite number of complex pairwise conjugated zeros, $z^2=-\omega ^2_k, \;\; z^2=-\overline{\omega }^2_k, \;\; k=1,2,...$\ and 
$\sum\limits_{k=1}^\infty  \omega ^{-2}_k $\ is absolutely convergent. \\
By virtue of the above properties one can write
\begin{eqnarray}
\frac{\omega ^4}{\Phi (z^2)}=\sum\limits_{i=1}^m \frac{\eta _i\Omega ^2_i}{1+\frac{z^2}{\Omega ^2_i}}+\sum\limits_{k=1}^\infty \left(\frac{\eta _k
\omega ^2_k}{1+\frac{z^2}{\omega ^2_k}}+\frac{\overline{\eta }_k\overline{\omega }^2_k}{1+\frac{z^2}{\overline{\omega }^2_k}} \right)  \label{w21}
\end{eqnarray}
Following \cite{b21} we define new variables
\begin{eqnarray}
&& \tilde Q_i\equiv \prod\limits_{\scriptstyle j=i \atop\scriptstyle j\neq i}^m 
\left(1+\frac{1}{\Omega ^2_j}\frac{d^2}{dt^2}\right)
\prod_{k=1}^\infty \left(1+\frac{1}{\omega ^2_k}\frac{d^2}
{dt^2}\right)\left(1+\frac{1}{\overline{\omega }^2_k}\frac{d^2}{dt^2}\right)q   \nonumber \\
&& Q_k\equiv \prod_{i=1}^m\left(1+\frac{1}{\Omega ^2_i}\frac{d^2}{dt^2}\right) \prod\limits_{\scriptstyle l=1 \atop\scriptstyle l\neq k }^\infty 
\left(1+\frac{1}{\omega ^2_l}\frac{d^2}{dt^2}\right)
\prod_{l=1}^\infty \left(1+\frac{1}{\overline{\omega }^2_l}\frac{d^2}{dt^2}\right)q     \label{w22} \\
&& \overline{Q}_k\equiv \prod_{i=1}^m\left(1+\frac{1}{\Omega ^2_i}\frac{d^2}{dt^2}\right)\prod_{l=1}^\infty \left(1+\frac{1}{\omega ^2_l}\frac{d^2}
{dt^2}\right)\prod\limits_{\scriptstyle l=1 \atop\scriptstyle l\neq k}^
\infty \left(1+\frac{1}{\overline{\omega }^2_l}\frac{d^2}{dt^2}\right)q   \nonumber
\end{eqnarray}
With the above definitions the Lagrangian takes the form
\begin{eqnarray}
&& L=-\frac{1}{2}\sum\limits_{i=1}^m \eta _i\tilde Q_i\left(\frac{d^2}{dt^2}+\Omega ^2_i\right)\tilde Q_i-  \label{w23}  \\
&& -\frac{1}{2}\sum\limits_{k=1}^\infty \left(\eta _kQ_k\left(\frac{d^2}{dt^2}+\omega ^2_k\right)Q_k+   
\overline{\eta }_k\overline{Q}_k\left(\frac{d^2}{dt^2}+\overline{\omega }^2_k\right)\overline{Q}_k\right)  \nonumber
\end{eqnarray}
By rescaling $\tilde Q_i\rightarrow \frac{\tilde Q_i}{\sqrt{\mid \eta _i\mid }}, \;\; Q_k\rightarrow \frac{Q_k}{\sqrt{\eta _k}}, \;\; \overline{Q}_k
\rightarrow \frac{\overline{Q}_k}{\sqrt{\overline{\eta }_k}}$, and passing to the Hamiltonian formalism we find
\begin{eqnarray}
H=\frac{1}{2}\sum\limits_{i=1}^m (sgn \;\eta _i)(\tilde P^2_i+\Omega ^2_i\tilde Q^2_i)+\frac{1}{2}\sum\limits_{k=1}^\infty  ((P^2_k+\omega ^2_kQ^2_k)+
(\overline{P}^2_k+\overline{\omega }^2_k\overline{Q}^2_k))    \label{w24}
\end{eqnarray}
This is, however, not the end of the story because the variables $Q_k, P_k$\ are complex. In order to find the relevant real variables we perform the 
following complex canonical transformation \cite{b21}
\begin{eqnarray}
&& P_k=\frac{1}{2}\sqrt{\omega _k}\left((p_{1k}+iq_{1k})+i(p_{2k}-iq_{2k})\right)  \nonumber \\
&& \overline{P}_k=\frac{1}{2}\sqrt{\overline{\omega }_k}((p_{1k}-iq_{1k})-i(p_{2k}+iq_{2k}))   \label{w25} \\
&& Q_k=\frac{i}{2\sqrt{\omega _k}}\left((p_{1k}-iq_{1k})+i(p_{2k}+iq_{2k})\right)   \nonumber \\
&& \overline{Q}_k=\frac{-i}{2\sqrt{\overline{\omega }_k}}\left((p_{1k}+iq_{1k})-i(p_{2k}-iq_{2k})\right)   \nonumber
\end{eqnarray}
which allows to rewrite eq. (\ref{w24}) as 
\begin{eqnarray}
&& H=  \label{w26}  \\
&& =\frac{1}{2}\sum\limits_{i=1}^m(sgn\;\eta _i)(\tilde P_i^2+\Omega ^2_i\tilde Q^2_i)-\sum\limits_{k=1}^\infty (Im\omega _k(q_{1k}p_{1k}+q_{2k}p_{2k})
+Re \omega _k(q_{1k}p_{2k}-q_{2k}p_{1k}))   \nonumber
\end{eqnarray}
So $H$\ is a sum (with alternating sign) of finite number of harmonic oscillators and an infinite number of terms which are linear combinations of 
dilatation and angular momentum in two dimensions. All terms depend on different variables and can be easily quantized; only dilatation calls for ordering 
rule - we adopt the simplest one: $q_1p_1+q_2p_2\rightarrow \frac{1}{2}(\hat q_1\hat p_1+\hat p_1\hat q_1+\hat q_2\hat p_2+\hat p_2\hat q_2)$.  \\
Let us consider one term of the second sum on the $RHS$\ of eq.(\ref{w26}). It is of the form
\begin{eqnarray}
\hat h=\frac{\mu }{2}(\hat q_1\hat p_1+\hat p_1\hat q_1+\hat q_2\hat p_2+\hat p_2\hat q_2)+\nu (\hat q_1\hat p_2-\hat q_2\hat p_1)   \label{w27}
\end{eqnarray}
Dilatations commute with angular momentum so the spectrum of $\hat h$\ reads
\begin{eqnarray}
\varepsilon _{n,\lambda }=\mu \lambda +\nu n, \;\;\; n=0,\pm 1,..., \;\;\;-\infty <\lambda <\infty     \label{w28}
\end{eqnarray}
and the corresponding wave functions (for example, in coordinate representation) can be easily found \cite{b21}.  \\
Now, let us consider classical equations of motion implied by $h$. They read
\begin{eqnarray}
\dot q_i=\mu q_i-\nu \varepsilon _{ik}q_k, \;\;\; \dot p_i=-\mu p_i+\nu \varepsilon _{ik}p_k  \label{w29}
\end{eqnarray}
The solutions are linear combinations of $exp ((\mu \pm i\nu )t)$\  (for $q's$\ ) or $exp((-\mu \pm i\nu )t)$\ (for $p's$\ ). Therefore, we are dealing with 
exponentially growing solutions. Following the same line of reasoning as in the first example one could conclude that the energy must take complex 
values. Again this conclusion results from the       assumption that the matrix elements of canonical variables in the energy representation are 
well-defined while it is straightforward to check that they are badly divergent (cf. eq.(34) of Ref. \cite{b21}). \\
Let us now make few remarks about path integral approach. Some care must be excercised when dealing with the Hamiltonians unbounded from below. Let 
us take as an example the inverted harmonic oscilator with the "frequency" $\omega $. As we have mentioned above the Hamiltonian is here a well-defined 
selfadjoint operator with purely continuous spectrum extending over the whole real line. \\
The propagator function
\begin{eqnarray}
K(x,y;t)=\sum\limits_{\lambda =\pm 1}\int\limits_{-\infty }^\infty  d\left(\frac{E}{\hbar \omega }\right)e^{\frac{-iEt}{\hbar }}
\Psi _{E\lambda }(x)\overline{\Psi }_{E\lambda }(y)    \label{w30} 
\end{eqnarray}
cannot be continued to imaginary time whatever the sign of $t$\ is. On the level of path integration the Euclidean path integral is not well-defined 
due to the fact that the integrand ceases to be of the form of the exponent of negative definite functional. \\
However, we can do the path integral directly (without refering to imaginary time approach) by applying Trotter formula and doing Fresnel integrals. 
The result, when extended to  whole time axis, reads
\begin{eqnarray}
K(x,y;t)=\frac{(1-isgn t)}{2\sqrt{\pi \hbar \mid t\mid }}\sqrt{\frac{\omega t}{sh \omega t}}e^{\frac{i\omega }{2\hbar sh\omega t}((x^2+y^2)
ch\omega t-2xy)}    \label{w31}
\end{eqnarray}
This result can be checked as follows. 
By virtue of eq. (\ref{w30}) one gets
\begin{eqnarray}
\int\limits_{-\infty }^\infty  dte^{\frac{i\tilde Et}{\hbar }}K(0,0;t)=\frac{2\pi }{\omega }\sum\limits_{\lambda =\pm 1}\mid \Psi _{\tilde E\lambda }
(0)\mid ^2  \label{w32}
\end{eqnarray}
Using eq. (\ref{w31}) and the explicit form of $\Psi _{\tilde E\lambda }$\ (cf. eq. (\ref{w11}) ) we find that (\ref{w32}) holds true. 

Consider now 
$K(0,0;t)$\ for $t>0$:
\begin{eqnarray}
K(0,0;t)=\frac{(1-i)}{2}\sqrt{\frac{\omega }{\pi \hbar sh\omega t}}    \label{w33}
\end{eqnarray}
Let us forget for a moment that the imaginary time representation is ill-defined. In order to extract the information about the spectrum of our 
Hamiltonian we continue (\ref{w33}) to imaginary time and consider the $t\rightarrow \infty $\ behaviour. We find that $K(0,0;t)$\ becomes periodic 
in Euclidean regime which means that the energies are imaginary! 

For our first example the propagator is simply the product of two propagators, one for harmonic oscillator and one for inverted harmonic oscillator. 
Therefore, we see that the inproper application of path integral method will result in wrong conclusions concerning the energy spectrum and unitarity. 

The same applies to the nonlocal harmonic oscillator. We found     that the Hamiltonian is a sum of commuting pieces, many of them having the spectrum 
unbounded from below. Again, it appears that an attempt to derive the energy eigenvalues from Euclidean version of path integral leads to complex 
energies, i.e. to the conclusion that the theory is not unitary.
\newpage
{\bf ACKNOWLEDGMENT} \\
P.K. thanks Barry Simon for kind and enlightening correspondence and bringing Ref. \cite{b17} to our attention.

\end{document}